\documentclass[aps,prl,twocolumn,superscriptaddress,showpacs]{revtex4}

\usepackage{graphicx}

\begin{document}

% Use the \preprint command to place your local institutional report
% number in the upper righthand corner of the title page in preprint mode.
% Multiple \preprint commands are allowed.
% Use the 'preprintnumbers' class option to override journal defaults
% to display numbers if necessary
%\preprint{}

%Title of paper
\title{Magnetic field symmetry and phase rigidity of the nonlinear conductance in a ring}

% repeat the \author .. \affiliation  etc. as needed
% \email, \thanks, \homepage, \altaffiliation all apply to the current
% author. Explanatory text should go in the []'s, actual e-mail
% address or url should go in the {}'s for \email and \homepage.
% Please use the appropriate macro foreach each type of information

% \affiliation command applies to all authors since the last
% \affiliation command. The \affiliation command should follow the
% other information
% \affiliation can be followed by \email, \homepage, \thanks as well.
\author{R. Leturcq}
\email[]{leturcq@phys.ethz.ch}
%\homepage[]{Your web page}
%\thanks{}
%\altaffiliation{}
\affiliation{Solid State Physics Laboratory, ETH Z\"urich, CH-8093 Z\"urich, Switzerland}

\author{D. S\'anchez}
\affiliation{Departament de F\'{\i}sica, Universitat de les Illes Balears, E-07122 Palma de Mallorca, Spain}

\author{G. G\"otz}
\affiliation{Solid State Physics Laboratory, ETH Z\"urich, CH-8093 Z\"urich, Switzerland}

\author{T. Ihn}
\affiliation{Solid State Physics Laboratory, ETH Z\"urich, CH-8093 Z\"urich, Switzerland}

\author{K. Ensslin}
\affiliation{Solid State Physics Laboratory, ETH Z\"urich, CH-8093 Z\"urich, Switzerland}

\author{D. C. Driscoll}
\affiliation{Materials Department, University of California, Santa Barbara, CA-93106, USA}

\author{A. C. Gossard}
\affiliation{Materials Department, University of California, Santa Barbara, CA-93106, USA}

%Collaboration name if desired (requires use of superscriptaddress
%option in \documentclass). \noaffiliation is required (may also be
%used with the \author command).
%\collaboration can be followed by \email, \homepage, \thanks as well.
%\collaboration{}
%\noaffiliation

\date{\today}

\begin{abstract}
We have performed nonlinear transport measurements as a function of a perpendicular magnetic field in a semiconductor Aharonov-Bohm ring connected to two leads. While the voltage-symmetric part of the conductance is symmetric in magnetic field, the voltage-antisymmetric part of the conductance is not symmetric. These symmetry relations are compatible with the scattering theory for nonlinear mesoscopic transport. The observed asymmetry can be tuned continuously by changing the gate voltages near the arms of the ring, showing that the phase of the nonlinear conductance in a two-terminal interferometer is not rigid, in contrast to the case for the linear conductance.
\end{abstract}

% insert suggested PACS numbers in braces on next line
\pacs{73.23.-b, 73.50.Fq, 73.63.-b}
% insert suggested keywords - APS authors don't need to do this
%\keywords{}

%\maketitle must follow title, authors, abstract, \pacs, and \keywords
\maketitle

% Introduction

A mesoscopic ring can be used as an electron interferometer in order to compare the electronic phase of electrons traveling through both arms of the ring using the Aharonov-Bohm (AB) effect. However, it has been shown that a two-terminal ring does not allow to measure directly this phase difference in the linear transport \cite{Yeyati02}: the two-terminal conductance shows AB oscillations with a phase constrained to 0 or $\pi$ \cite{GefenButtiker,Yacoby01,Yacoby03}. This phase rigidity is a consequence of microreversibility \cite{OnsagerCasimir} showing that the linear conductance of a two-terminal system must be symmetric in magnetic field \cite{Buttiker01,Yeyati02}. A direct measurement of the phase difference is possible only in an open multi-terminal geometry \cite{Yacoby04,SchusterSigrist}.

While the Onsager-Casimir relations hold close to equilibrium (linear conductance), there is no fundamental reason why far from equilibrium the nonlinear conductance should still follow this symmetry, i.e., one could expect $G(V,B) \neq G(V,-B)$. It is then natural to ask whether the phase rigidity would still hold for the nonlinear transport in a two-terminal ring.

In a phase coherent diffusive system, nonlinear conductance is expected when the bias voltage is larger than $E_T/e$, where $E_T$ is the Thouless energy \cite{AltshulerLarkinKhmelnitskii}. Models developed for non-interacting electrons predict an effect symmetric in magnetic field, which has been observed experimentally through bias voltage induced universal conductance fluctuations \cite{WebbVegvar,Kaplanandother}. The possibility to observe magnetic field asymmetric nonlinear transport has been addressed only very recently both theoretically \cite{Sanchez03,Spivak05} and experimentally \cite{Lofgren01,WeiJ01,Zumbuhl04,Marlow01}. The models proposed there rely on effects of electron-electron interactions in noncentrosymmetric systems. Such behavior could be also expected in AB rings, for which symmetry breaking occurs due to asymmetries in the phase accumulated in each arm of the ring.

Here we address the question of the magnetic field symmetries of the nonlinear conductance in a ring used as an Aharonov-Bohm interferometer. We have performed nonlinear d.c. transport measurements in a ring connected to two terminals. The current is fitted by a polynomial function of the bias voltage, with each coefficient of the decomposition showing AB oscillations as a function of magnetic field. While the odd 
coefficients are symmetric in magnetic field and show strong $h/2e$ oscillations, the even coefficients are asymmetric in magnetic field with weak $h/2e$ oscillations. Furthermore, an electrostatic change of the phase in one arm of the ring produces a continuous change of the phase of the asymmetric coefficients, suggesting that the asymmetry is related to the electronic phase. The symmetry relations can be understood within a simple model using scattering theory. This simple model, however, cannot explain the weak amplitude of $h/2e$ oscillations in the even coefficients.

% Experimental setup

\begin{figure}[b]
\includegraphics[width=1\linewidth]{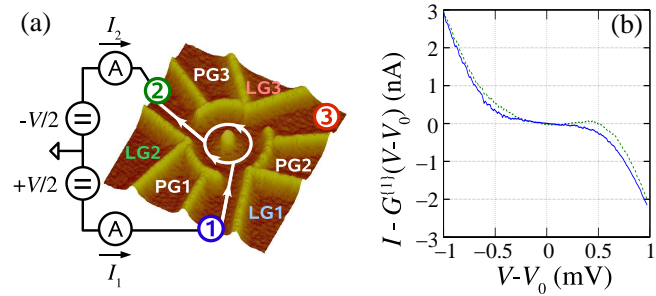}%
\caption{\label{fig1} (Color online) (a) AFM micrograph of the ring and scheme of the experimental setup for the nonlinear measurement. The electron paths are sketched by the white lines around the ring. (b) Nonlinear part of the current-voltage characteristics taken by applying a symmetric bias voltage $V$ between leads 1 and 2 and measuring the current through lead 1. The nonlinear part is obtained after subtracting from the current $I$ the linear part of the fit, i.e., $G^{\{1\}}(V-V_0)$. The gate voltages are  $V_{\text{PG2}} = V_{\text{PG3}} = 0$ and $V_{\text{PG1}}=-0.050$ V. The traces are taken at two magnetic fields of approximately the same amplitude but opposite signs, $B=-0.0583$ T (dashed line) and $B=+0.0570$ T (solid line).}
\end{figure}

The sample has been fabricated on a GaAs/GaAlAs heterostructure containing a two-dimensional electron gas (2DEG) 34 nm below the surface. A back-gate has been used to tune the electron density to $4.5 \times 10^{15}$ m$^{-2}$ and a mobility of 27 m$^2$(Vs)$^{-1}$. The surface of the heterostructure is patterned by local oxidation with an atomic force microscope (AFM), defining depleted regions in the 2DEG below the oxide lines \cite{HeldFuhrer}. Figure~\ref{fig1}(a) shows an AFM image of the ring, which is initially connected to three terminals (labelled 1 to 3) through quantum point contacts (QPCs). The opening of the QPCs can be controlled by the three gates labelled LG1 to LG3, while the gates PG1 to PG3 control the electron density in each arm of the ring. In this experiment, the QPC connected to lead 3 is pinched off by applying a negative voltage on LG3, in order to perform an effective \emph{two-terminal measurement}. This is checked by measuring the current flowing through lead 3, which is below 10 pA for all measurements, and does not depend on the bias voltage. All measurements have been performed at 1.7 K in a $^4$He cryostat with a variable temperature insert.

The d.c. $I-V$ characteristics are measured by applying voltages $V_1=+V/2$ and $V_2=-V/2$ on leads 1 and 2, through two identical circuits, including $I/V$ converters for the current measurement [Fig.~\ref{fig1}(a)]. This configuration allows to minimize circuit induced asymmetries \cite{Lofgren01}, and we have checked carefully that interchanging leads 1 and 2 in the setup gives the same results.

% I-V trace and decomposition of the conductance

The nonlinear part of a typical $I-V$ characteristic is shown in Fig.~\ref{fig1}(b), for a particular setting of the gate voltages and for two magnetic fields of the same amplitude but opposite signs. The $I-V$ curves show a clear nonlinear behavior at voltages above about 100 $\mu$V. Furthermore, they demonstrate that the conductance is not exactly equal for opposite magnetic fields.

In order to quantify the nonlinear behavior, the $I-V$ curve is fitted with a fifth order polynomial \cite{NotePolynomial}, allowing a voltage offset $V_0$ \cite{NoteOffset}:
\begin{equation}
I = \sum_{n=1}^5 G^{\{n\}} (V-V_0)^n.
\end{equation}
Here $G^{\{n\}}$ and $V_0$ are fitting parameters. In this decomposition, the odd coefficients ($G^{\{1\}}$, $G^{\{3\}}$,...) correspond to the voltage-symmetric part of the differential conductance $G(V,B)=dI/dV$, and the even coefficients ($G^{\{2\}}$, $G^{\{4\}}$,...) to the voltage-antisymmetric part of $G(V,B)$.

\begin{figure}[b]
\includegraphics[width=1\linewidth]{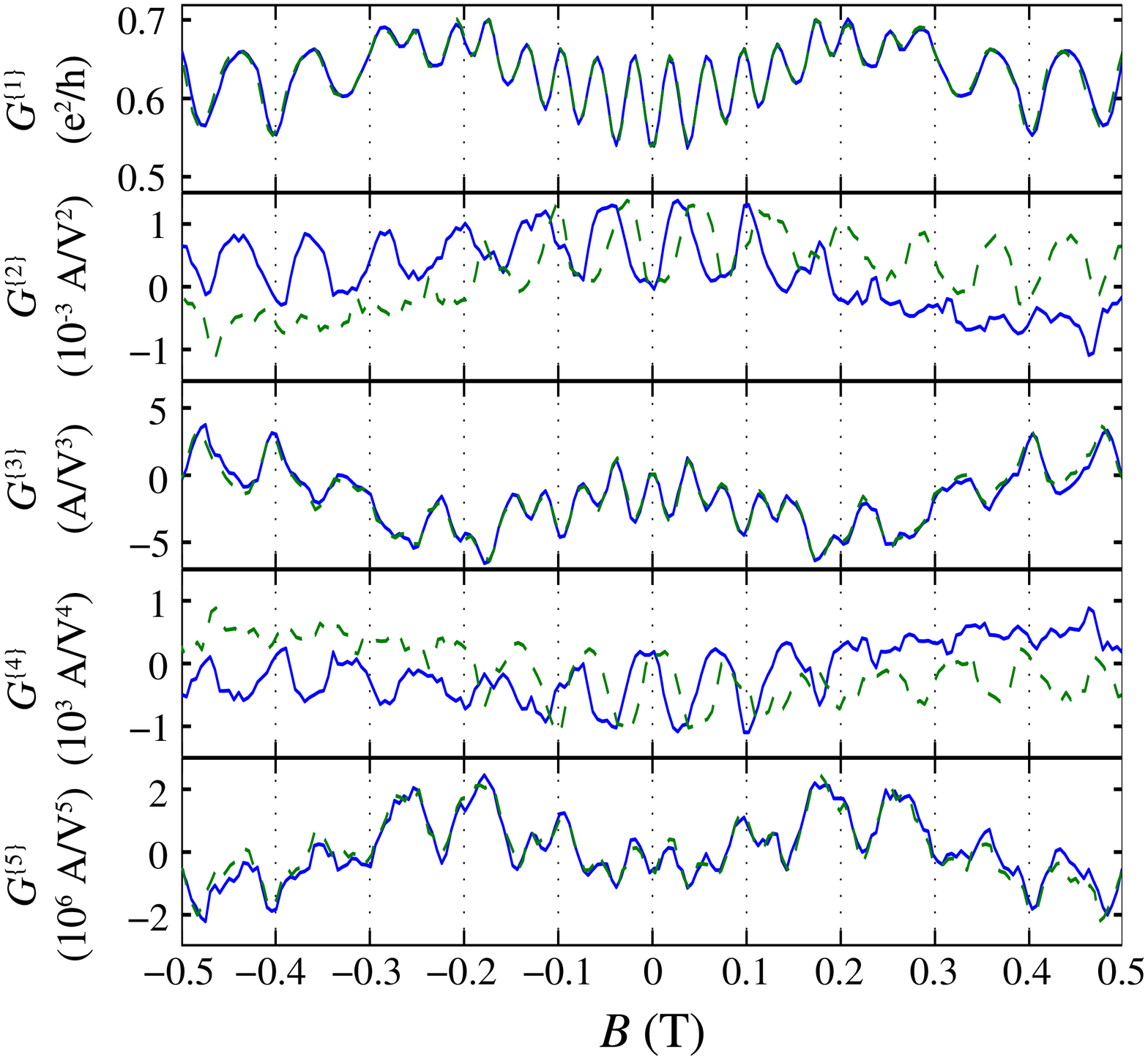}%
\caption{\label{fig2} (Color online) Magnetic field dependence of the nonlinear conductance coefficients $G^{\{n\}}$, for $V_{PG2} = V_{PG3} = 0$ and $V_{PG1}=-0.050$ V. The dashed curves correspond to the inversion of the plain curves compared to $B=0$. An offset of the magnetic field of 0.2 mT has been corrected.}
\end{figure}

The fitting parameters are shown as a function of magnetic field in Fig.~\ref{fig2} for fixed gate voltages. In each panel, the dashed line corresponds to the same curve as the plain line, but mirrored horizontally at $B=0$ in order to check the magnetic field symmetry. All conductances show AB oscillations as a function of magnetic field, with a period close to 75 mT. This period corresponds to a diameter of the ring of 260 nm, compatible with the lithographic size of the ring. In addition to $h/e$ oscillations, the odd coefficients show strong $h/2e$ oscillations. It is then interesting to note that the even coefficients do not show any significant $h/2e$ oscillations.

It is clear from the uppermost panel that the linear conductance $G^{\{1\}}$ is symmetric in magnetic field. The higher order coefficients show a remarkable behavior. While the odd coefficients are symmetric in magnetic field within experimental errors, the even coefficients are not symmetric in magnetic field.

% Change asymmetry with a gate voltage

\begin{figure}[b]
\includegraphics[width=1\linewidth]{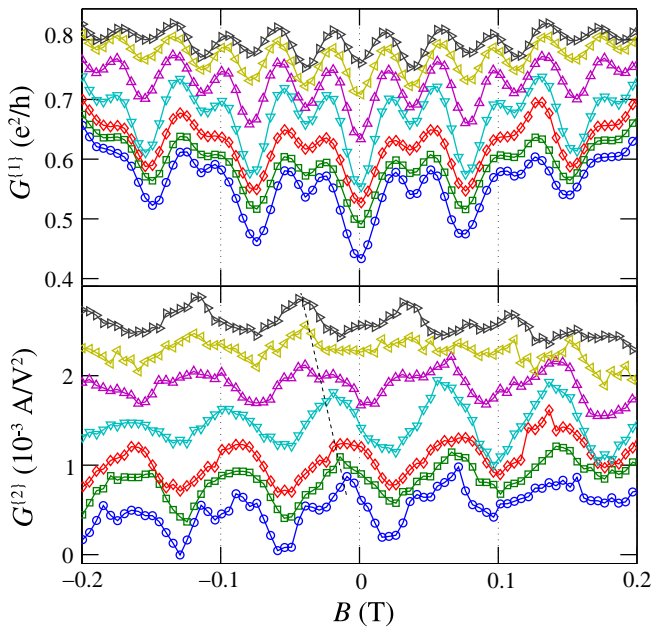}%
\caption{\label{fig3} (Color online) First and second order conductances vs. magnetic field, for several gate voltages $V_{\text{PG2}} = V_{\text{PG3}}$, varying from $+0.125$ V (bottom curves) to $-0.025$ V (top curves), by steps of $0.025$ V. The three upper curves for $G^{\{1\}}$ have been shifted vertically by respectively $0.077$, $0.155$ and $0.232 \times e^2/h$ (from bottom to top). The curves for $G^{\{2\}}$ have been shifted vertically by respectively $4$, $8$, $12$, $16$, $20$, $28$ and $32 \times 10^{-4}$ A/V$^2$ (from bottom to top). The dashed line in the lower panel points out the shift of one maximum.}
\end{figure}

In order to investigate the asymmetry of the even coefficients further, we have used lateral gates PG1, PG2 and PG3 [Fig.~\ref{fig1}(a)] to modify locally the electron density in the ring. Earlier experiments have shown that a gate voltage can tune the electronic phase accumulated along the arms of the ring through the product $k_F L$, where $k_F$ is the Fermi wave vector and $L$ the length of the ring affected by the gate \cite{Vegvar01,Cernicchiaro01,Pedersen01,Yacoby04,Yacoby03}. Here we use either the gates PG2 and PG3 (right arm of the ring), or the gate PG1 (left arm of the ring), in order to change the phase only in a selected part of the ring. The two first order coefficients $G^{\{1\}}$ and $G^{\{2\}}$ are shown in Fig.~\ref{fig3} for several gate voltages $V_{\text{PG2}} = V_{\text{PG3}}$ and fixed $V_{\text{PG1}}=0$. The higher order coefficients show identical results, namely odd coefficients behave comparably to $G^{\{1\}}$ and even coefficients to $G^{\{2\}}$. The change in $G^{\{1\}}$ is similar to earlier two-terminal experiments, where the relative amplitude between $h/e$ and $h/2e$ oscillations is tuned due to a change of the phase accumulated along the ring \cite{Pedersen01,Yacoby03}. The surprising result is that the phase of the oscillations of $G^{\{2\}}$ changes continuously with the gate voltage. Up to now, such continuous phase shifts have been reported only in the linear conductance of multi-terminal interferometers, where they are directly related to a change in the phase difference between both arms of the ring \cite{Yacoby04,SchusterSigrist}. A similar result has been obtained by sweeping the gate voltage $V_{\text{PG1}}$, keeping $V_{\text{PG2}} = V_{\text{PG3}}=0$.

\begin{figure}[b]
\includegraphics[width=1\linewidth]{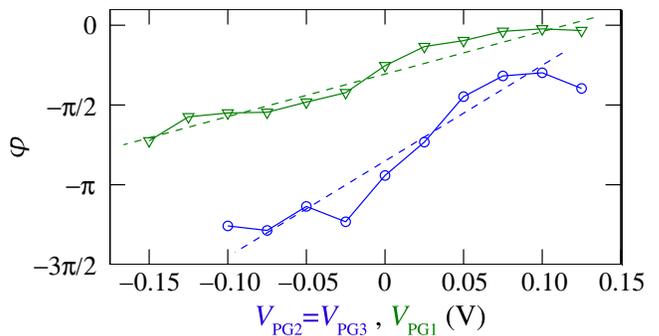}%
\caption{\label{fig4} (Color online) Phase of the magnetic field oscillations of $G^{\{2\}}$ as a function of $V_{\text{PG2}} = V_{\text{PG3}}$ for $V_{\text{PG1}}=0$ (circles), and as a function of $V_{\text{PG1}}$ for $V_{\text{PG2}} = V_{\text{PG3}} = 0$ (triangles). The dashed lines are guides to the eye, showing a linear dependence with slope $0.85 \pi$/V (top) and  $1.90 \pi$/V (bottom).}
\end{figure}

We have evaluated the phase of the oscillating part of $G^{\{2\}}$, $\varphi$, with the formula $\tan ( \varphi ) = \left< G^{\{2\}}(B) \sin \left( B/B_0 \right) \right> / \left< G^{\{2\}}(B) \cos \left( B/B_0 \right) \right>$, where $B_0$ is the AB period and $\left< ... \right>$ is the mean taken over the full magnetic field range. Figure~\ref{fig4} shows the variation of the phase for both gate sweeps. Both slopes differ by a factor close to 2, as expected from the lever arms of these gates, but surprisingly they have the same sign. A possible explanation is that the phase is not controlled locally on each arm of the ring, but that each gate rather affects the electron density in the whole ring.

% Origin of the non-linear conductance

We now discuss the possible origin of the nonlinear conductance. While electron heating could explain both the origin and the symmetry of the odd coefficients, it cannot explain the even ones, since the temperature should depend on the electric power. We have checked that nonlinearities on the QPCs depend only weakly on the magnetic field. In addition, the different magnetic field symmetry observed for the odd and even coefficients excludes thermoelectric effects \cite{MolenkampGodijn} or spurious circuit-induced nonlinear effects, that would give the same symmetry for all terms.

Previous nonlinear transport experiments in AB rings \cite{WebbVegvar} were explained in terms of scattering theory with a bias voltage dependent transmission \cite{AltshulerLarkinKhmelnitskii}. The transmission is expected to be modified due to a global shift in energy of electronic paths in the ring induced by the finite bias voltage. An additional effect could come from the electrostatic Aharonov-Bohm effect, for which an electrostatic potential along the electronic paths modifies the electronic phase \cite{OudenaardenVdWiel}. In our experiment, we are not able to distinguish both origins of nonlinearities.

% Origin of the asymmetry

Theoretical studies have shown that the $I$--$V$ characteristics of a mesoscopic system are, quite generally, not even under reversal of the magnetic field because the response of the screening potential is $B$-asymmetric at large bias \cite{Sanchez03,Spivak05}. Such interaction-induced asymmetry was not reported in previous nonlinear measurements on rings \cite{WebbVegvar}, due to the multi-terminal nature of these experiments, but it has been observed very recently in carbon nanotubes \cite{WeiJ01} and in semiconductor quantum dots \cite{Zumbuhl04,Marlow01}. These experiments address the lowest-order nonlinearity $G^{\{2\}}$ only. Hence, our observation that the even (odd) coefficients $G^{\{n\}}$ are asymmetric (symmetric) in magnetic field is novel. We now demonstrate that this effect follows from general arguments.

The scattering theory for nonlinear mesoscopic transport \cite{Christen02} establishes that the current,
\begin{equation}\label{eq_cur}
I=\frac{2e}{h} \int dE \, T(E;U) [f(E-eV/2)-f(E+eV/2)]\,,
\end{equation}
depends on the transmission $T$, which is a functional of the screening potential $U$ in the mesoscopic conductor. For simplicity, we assume that $U$ is uniform though the full theory takes into account the spatial distribution. In Eq.~(\ref{eq_cur}), $f(x)=1/(1+\exp{(x-E_F)/k_B T})$ with $E_F$ the Fermi energy. Within the Fermi-Thomas approximation, the induced charge density is a linear function of the external bias $V$. Assuming that the density of states is weakly energy dependent, charge conservation demands that $U(V)=U_{\rm eq} + u V$, where the characteristic potential $u=(\partial U/\partial V)_{\rm eq}$ relates the change in the screening due to a voltage shift. Such response is, in general, asymmetric under $B$ reversal \cite{Sanchez03}. We emphasize that this effect arises in the nonlinear regime only since microreversibility requires $U_{\rm eq}(B)=U_{\rm eq}(-B)$ at equilibrium.

Inserting $U(V)$ in Eq.~(\ref{eq_cur}) and expanding in powers of $V$, we find at zero temperature:
\begin{equation}
G^{\{n\}}=\frac{1}{2^{n-2}n!} \frac{e^{n+1}}{h} T^{(n-1)}(E_F)
[\delta_{n,{\rm odd}}-2u\delta_{n,{\rm even}}]\,.
\end{equation}
Clearly, since the transmission and its energy derivatives $T^{(n)}$ are $B$-symmetric, only the even coefficients are asymmetric under $B$ reversal. The internal potential depends on the phase accumulated along both arms, which can be tuned with gate voltages, thus affecting the magnetic field asymmetry. This asymmetry arises when the conductor is asymmetrically coupled to the leads (via scattering or capacitive coupling \cite{Sanchez03}). It is very likely that our ring is geometrically asymmetric, due to different sizes of the arms [see Fig.~\ref{fig1}(a)] or to randomly distributed defects.

% Suppression of h/2e oscillations in the nonlinear conductance

We note however that this scenario does not explain the experimentally observed weak amplitude of $h/2e$ oscillations in the even coefficients. While $h/e$ oscillations are due to interference effects between paths going once around the ring, $h/2e$ oscillations have several possible origins. They can be due to interference between paths going twice around the ring \cite{GefenButtiker}, or to interference between time-reversed paths going each once around the ring (Al'tshuler, Aronov and Spivak (AAS) effect \cite{Altshuler01}). Interestingly, the last mechanism is not sensitive to the electric phase accumulated by the electron along the ring, since both time-reversed paths will accumulate the same phase. The fact that we can influence the ratio of $h/e$ and $h/2e$ oscillations in the linear conductance using a gate [see Fig. \ref{fig3}] shows that both effects contribute to $h/2e$ oscillations.

% Conclusion

In conclusion, we have made nonlinear transport measurements in a two-terminal ring. The nonlinear conductance shows AB oscillations as a function of magnetic field. We show that the voltage-symmetric part of the conductance is symmetric in magnetic field, while the voltage-antisymmetric part is asymmetric in magnetic field, compatible with the scattering theory for nonlinear transport. Furthermore we can tune the phase of the asymmetry by changing the voltage of gates placed nearby the ring, which shows that the nonlinear conductance is not phase-rigid in contrast to the linear conductance.

\begin{acknowledgments}
We thank M. B\"uttiker and M. Polianski for useful discussions. We acknowledge support from the Swiss Science Foundation (Schweizerischer Nationalfonds) via NCCR Nanoscience, the EU Human Potential Program via the Bundesministerium f{\"u}r Bildung und Wissenschaft, and the Spanish "Ram\'on y Cajal" program.
\end{acknowledgments}

% Create the reference section using BibTeX:
\bibliography{biblio}

\begin{thebibliography}{33}
\bibitem{NotePolynomial}
In the experiments presented here, the bias voltage range is -1 to +1 mV, and we have used a fifth order polynomial. We have checked that changing the order or the range does not change qualitatively the results.
\bibitem{NoteOffset}
A constant offset voltage, $V_0$, of the order of 10 $\mu$V is expected in our setup due to offset voltages of the amplifiers and thermal voltages along the cables. In addition, we observe AB oscillations on the offset voltage, which might come from rectification of an a.c. signal by the even terms of the conductance. This parasitic a.c. signal could be due to RF noise. However, we have checked that this offset does not change significantly the determination of the nonlinear conductance.
\end{thebibliography}

\end{document}